\documentclass[prl,
intlimits,sumlimits,
twocolumn,
showpacs,
showkeys,
]{revtex4}

\usepackage{dsfont}
\usepackage{amsmath}
\usepackage{subfig}
\usepackage{graphicx}
\usepackage{color}

\begin{document}
\title{Non--Stationarity in Financial Time Series and Generic Features}
\author{Thilo A. Schmitt}
\email{thilo.schmitt@uni-due.de}
\affiliation{Fakult\"at f\"ur Physik, Universit\"at Duisburg--Essen, Duisburg, Germany}
\author{Desislava Chetalova}
\affiliation{Fakult\"at f\"ur Physik, Universit\"at Duisburg--Essen, Duisburg, Germany}
\author{Rudi Sch\"afer}
\affiliation{Fakult\"at f\"ur Physik, Universit\"at Duisburg--Essen, Duisburg, Germany}
\author{Thomas Guhr}
\affiliation{Fakult\"at f\"ur Physik, Universit\"at Duisburg--Essen, Duisburg, Germany}

\date{\today}

\begin{abstract}
  Financial markets are prominent examples for highly non--stationary
  systems. Sample averaged observables such as variances and
  correlation coefficients strongly depend on the time window in which
  they are evaluated.  This implies severe limitations for approaches
  in the spirit of standard equilibrium statistical mechanics and
  thermodynamics.  Nevertheless, we show that there are similar
  generic features which we uncover in the empirical return
  distributions for whole markets. We explain our findings by setting
  up a random matrix model.
\end{abstract}



\maketitle

The great success of statistical mechanics and thermodynamics is borne
out by their ability to characterize, in the equilibrium, large
systems with many degrees of freedom in terms of a few state
variables, for example temperature and pressure. Ergodicity (or
quasi--ergodicity) is the prerequisite needed to introduce statistical
ensembles.  Systems out of equilibrium or, more generally,
non--stationary systems still pose fundamental challenges
\cite{Gao1999,Hegger2000,Bernaola-Galvan2001,Rieke2002}. Complex
systems --- the term ``complex'' is used in a broad sense --- show a
wealth of different aspects which can be traced back to
non--stationarity \cite{Zia2004,Zia2006}. Financial markets are
presently in the focus, because they demonstrated their
non--stationarity in a rather drastic way during the recent years.  To
assess a financial market as a whole, the correlations between the
prices of the individual stocks are of crucial importance
\cite{Zhang2011,Song2011,Munnix2012,Sandoval2012}.  They fluctuate
considerably in time, \textit{e.g.}, because the market expectations
of the traders change, the business relations between the companies
change, particularly in a state of crisis, and so on.  The motion of
the stock prices is in this respect reminiscent of that of particles
in many--body systems such as heavier atomic nuclei.  Depending on the
excitation energy, the motion of the individual particles can be
incoherent, \textit{i.e.}, uncorrelated in the above terminology, or
coherent (collective), \textit{i.e.}, correlated, or even somewhere
in--between~\cite{bohr1969nuclear,zelevinsky1996quantum,Guhr1998}.
This non--stationarity on the energy scale leads to very different spectral
properties, \cite{zelevinsky1996quantum,Guhr1998}. Such an analogy can be helpful,
but we do not want to overstretch it.

Here, we want to show that the non--stationarity, namely the
fluctuation of the correlations, induces generic features in financial
time series. These become visible when looking at quantities which
measure the stock price changes for the entire market. We have four
goals. First, we carry out a detailed data analysis revealing the
generic features.  Second, we set up a random matrix model to explain
them. Third, we demonstrate that the non--stationarity of the
correlations leads to heavy tails.  Fourth, we argue that our approach
maps a non--invariant situation to an effectively invariant one. For
an economic audience we discuss the consequences for portfolio
management elsewhere~\cite{Chetalova2013}.

\begin{figure}[htbp]
  \begin{center}
    \includegraphics[width=0.45\textwidth]{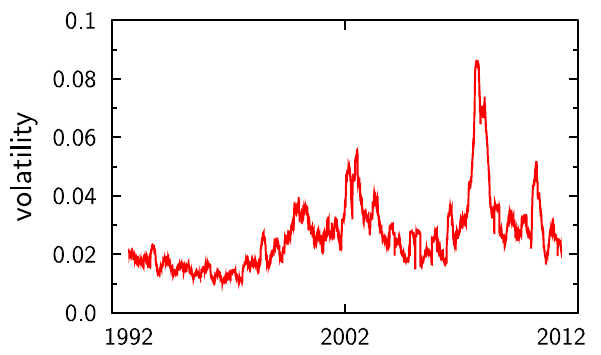}
  \end{center}
  \caption{Volatility time series for Goodyear from 1992 to 2012.
    The return interval is $\Delta t=1$ trading day and the time window has length
    $T= 60$ trading days.}
 \label{fig1}
\end{figure}

Consider $K$ companies with stock prices $S_k(t), \ k=1,\ldots,K$ as
functions of time $t$. The relative price changes over a fixed
time interval $\Delta t$, \textit{i.e.}, the returns
\begin{equation}
 r_k (t) = \frac{S_k(t + \Delta t)-S_k(t)}{S_k(t)}
\label{returns}
\end{equation}
are well--known to have distributions with heavy tails, the smaller
$\Delta t$, the heavier.  This is linked to the also well established
fact that the sample standard deviations $\sigma_k$, referred to as
volatilities, strongly fluctuate for different time windows of the
same length $T$ \cite{Black:1976fj,Schwert1989}, as shown in
Fig.~\ref{fig1}. This non--stationarity is not grasped by the
conventional
description of stock prices using Brownian motion
\cite{Bachelier:1900kx}, because the latter assumes a constant
volatility. Other schematic processes such as the (generalized)
autoregressive conditional heteroscedasticity models
\cite{Engle1982,Bollerslev1986} treat the volatilities as a stochastic
variable as well and can describe the heavy tails, but their
parameters lack a clear economic interpretation. This is not too
surprising, as recent studies \cite{Farmer2004,Schmitt2012} show that
the heavy tails result, among other reasons, from the very procedure
of how stock market trading takes place. 

\begin{figure}[htbp]
  \begin{center}
    \includegraphics[width=0.235\textwidth]{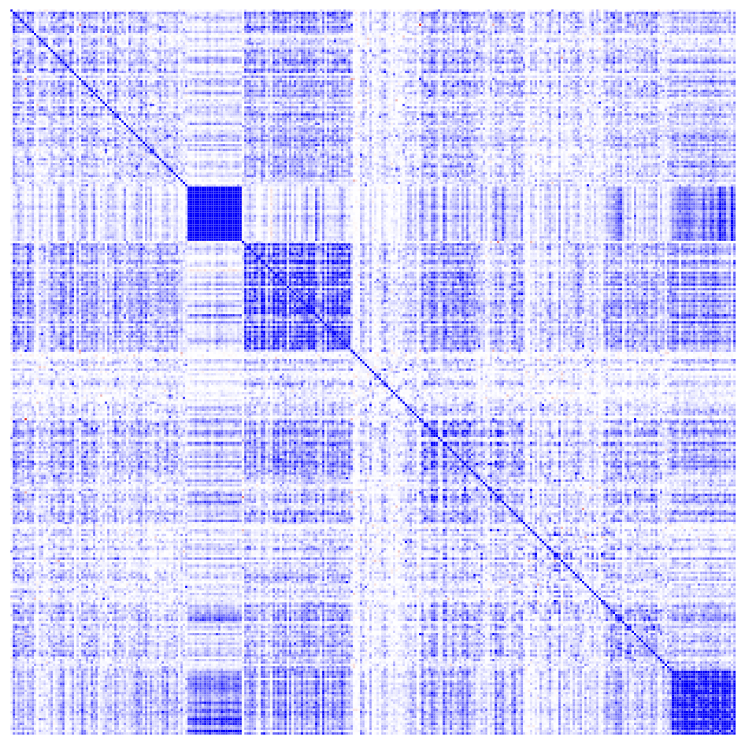}
    \includegraphics[width=0.235\textwidth]{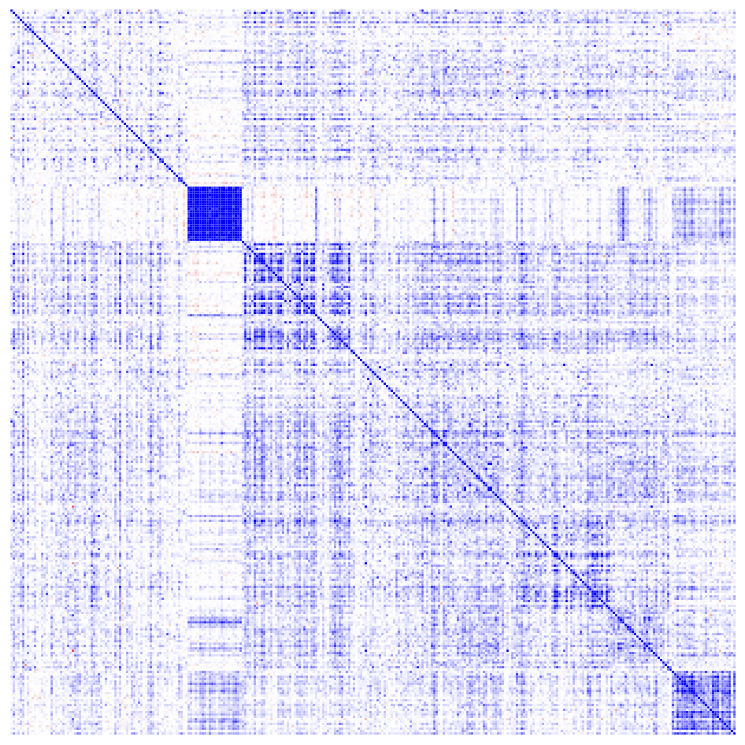}
  \end{center}
  \caption{Correlation matrices of $K=306$ companies for the fourth
    quarter of 2005 and the first quarter of 2006, the darker, the
    stronger the correlation. The companies are sorted according to
    industrial sectors.}
 \label{fig3}
\end{figure}

Another equally fundamental non--stationarity of stock market data
is the time dependent fluctuation of the sampled Pearson correlation
coefficients
 \begin{eqnarray}
C_{kl} &=& \left\langle  M_k(t) M_l(t) \right\rangle_T \nonumber\\
 M_k(t) &=& \frac{r_k(t) - \langle r_k(t)\rangle_T}
                                              {\sigma_k}
\label{corr}
\end{eqnarray}
between the two companies $k$ and $l$ in the time window of length
$T$, where $\sigma_k$ is evaluated in the same time window. The time
series $M_k(t)$ are normalized to zero mean and unit variance.  To
illustrate how strongly the $K\times K$ correlation matrix $C$ as a
whole changes in time, we show it for subsequent time windows in
Fig.~\ref{fig3}.
The dataset used here and in all analyses in the sequel consists of
$K=306$ continuously traded companies in the SP500 index between
1992 and 2012~\cite{yahoo}.  For later discussion, we emphasize that the
stripes in these correlation matrices indicate the structuring of the
market in industrial sectors, see, \text{e.g.},
Ref.~\cite{Munnix2012}. We will also use the covariance matrix $\Sigma
= \sigma C \sigma$ where the diagonal matrix $\sigma$ contains the
volatilities $\sigma_k, k=1,\ldots,K$.

\begin{figure}[htbp]
  \begin{center}
    \includegraphics[width=0.45\textwidth]{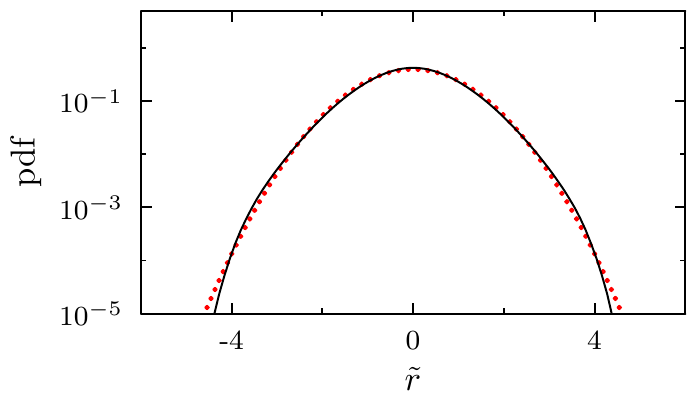}
  \end{center}
  \caption{Aggregated distribution of returns, here denoted by
    $\tilde{r}$, for fixed covariances from the SP500 dataset,
    $\Delta t=1$ trading day and window length $T=25$ trading days. The circles show a normal
    distribution.}
 \label{fig4}
\end{figure}

We now show that the returns are to a good approximation multivariate
Gaussian distributed, if the covariance matrix $\Sigma$ is fixed. Hence, we
assume that the distribution of the $K$ dimensional vectors
$r(t)=(r_1(t),\ldots,r_K(t))$ for a fixed return interval $\Delta t$
while $t$ is running through the dataset is given by
\begin{equation}
g(r|\Sigma)  = \frac{1}{\sqrt{\det(2\pi\Sigma)}} 
                                      \exp\left( -\frac{1}{2} r^\dagger \Sigma^{-1}r\right)  \ ,
\label{multivar}
\end{equation}
where we suppress the argument $t$ of $r$ in our notation. We test
this assumption with the SP500 dataset. We divide the time series in
windows of length $T$ (not to be confused with the return intervals of length $\Delta t$) which are so
short that the sampled covariances can be viewed as constant within
these windows.  However, the corresponding covariance matrices
$\Sigma$ are non--invertible because their rank is lower than $K$.
Mathematically, this is not a problem, because the distribution
\eqref{multivar} is still well--defined in terms of proper $\delta$
functions. To carry out the data analysis, we take all pairs $r_k,r_l$
of returns which, according to our assumption, should be bivariate
Gaussian distributed with a $2\times 2$ covariance matrix
$\Sigma^{(k,l)}$, which always is invertible. We rotate the
two--component vectors $(r_k,r_l)$ into the eigenbasis of
$\Sigma^{(k,l)}$, and normalize the axes with the eigenvalues.  All
distributions obtained in this way are aggregated.  Figure~\ref{fig4}
shows the results for daily returns and a window length of $T=25$
trading days. We observe a good agreement with a Gaussian.

However, as argued above, the covariances change in time, \textit{i.e.}, the
covariance matrix is not constant. Thus there is no contradiction to
the heavy--tailed distributions observed for \textit{individual}
stocks over longer time horizons.  Our idea is now to take the
non--stationarity of the covariance matrix into account by replacing
the fixed covariance matrix with a random matrix,
\begin{equation}
\Sigma \quad \longrightarrow \quad \frac{1}{N}AA^\dagger \ ,
\label{random}
\end{equation}
with $A$ being a rectangular $K\times N$ real random matrix without
any symmetries and with the dagger indicating the transpose.  The
product form $AA^\dagger/N$ is essential to model a proper covariance
matrix.  This is seen from the definition \eqref{corr} implying that
$C=MM^\dagger/T$ where the rectangular $K\times T$ data matrix $M$
contains the normalized time series $M_k(t)$ as rows.  Hence, when
viewing the rows of $A$ as model time series, their length is $N$, not
$T$. This important ingredient of our model and the meaning of $N$,
which is arbitrary at the moment, will be discussed later on. What is
the probability distribution of the random matrices $A$? --- Following
Wishart \cite{Wishart1928,muirhead2009aspects}, we assume the Gaussian
\begin{equation}
w(A|\Sigma) =  \frac{1}{\det^{N/2}(2\pi \Sigma)} 
                  \exp\left(-\frac{1}{2} \text{tr} A^\dagger \Sigma^{-1} A \right) \ , 
\label{wishart}
\end{equation}
where $\Sigma$ is now the empirical covariance matrix evaluated over
the entire time interval, \textit{i.e.}, it is fixed. The Wishart
covariance matrices $AA^\dagger/N$ fluctuate around the sampled
empirical one.  By definition, one has $\langle AA^\dagger \rangle/N =
\Sigma$, where the angular brackets without an index $T$ always
indicate the average over the ensemble defined by the distribution
\eqref{wishart}.  The parameter $N$ acquires the meaning of an inverse
variance characterizing the fluctuations around $\Sigma$. The larger
$N$, the more terms contribute to the individual matrix elements of
$AA^\dagger/N$, eventually making them sharp for $N\to\infty$.

In our model, the fluctuating covariances alter the multivariate
Gaussian \eqref{multivar}, implying the introduction of an ensemble
averaged return distribution
\begin{equation}
\langle g\rangle(r|\Sigma,N) = \int g\left(r\bigg|\frac{1}{N}AA^\dagger\right)
                                                                      w(A|\Sigma) d[A] \ ,
\label{mvaver}
\end{equation}
which parametrically depends on the fixed empirical covariance matrix
$\Sigma$ as well as on $N$.  The construction \eqref{mvaver} states
the most important conceptual point of our study. The measure $d[A]$
is simply the product of all differentials.  To calculate this
ensemble averaged return distribution, we write the Gaussian
\eqref{multivar} as a Fourier transform in terms of a $K$ component
vector $\omega$. The ensemble average is then Gaussian, leading to
\begin{eqnarray}
\langle  g \rangle (r|\Sigma,N) &=&  \frac{1}{\det^{N/2}(2\pi\Sigma)} 
                                                   \int \frac{d[\omega]}{(2\pi)^K} \exp(- \text{i} \omega\cdot r) 
                                                                                                              \nonumber\\
& &\quad   \int d[A] \exp\left(-\frac{1}{2} \text{tr} A^\dagger \Sigma^{-1} A \right) 
                                                                                                              \nonumber\\
& &\qquad   \exp\left(-\frac{1}{2N}\omega^\dagger AA^\dagger\omega\right) 
                                                                                                              \nonumber\\
&=&\frac{1}{(2\pi)^K} \int \frac{\exp( -\text{i} \omega\cdot r) d[\omega]}
                             {\det^{N/2}(1_K+\Sigma\omega\omega^\dagger/N)} \ .
\label{r1}
\end{eqnarray}
The matrix adding to the $K\times K$ unit matrix $1_K$ in the
determinant has rank unity, implying that the whole determinant is
equal to the positive definite quantity
$1+\omega^\dagger\Sigma\omega/N$.  Furthermore, using the expression
\begin{equation}
\frac{1}{a^\eta} = \frac{1}{2^\eta\Gamma(\eta)} \int_0^\infty z^{\eta-1} \exp\left(-\frac{a}{2}z\right) dz
\label{gamma}
\end{equation}
for real and positive variables $a$ and $\eta$, we find
\begin{eqnarray}
\langle  g \rangle (r|\Sigma,N) &=& \frac{1}{2^{N/2}\Gamma(N/2)} 
                                                 \int\limits_0^\infty dz \ z^{N/2-1} \exp\left(-\frac{z}{2}\right)
                                                                                       \nonumber\\
& & \int \frac{d[\omega]}{(2 \pi)^K} 
                   \exp \left(- \text{i} \omega \cdot r - \frac{z}{2N} \omega^\dagger\Sigma\omega\right)
\label{r2}
\end{eqnarray}
The $\omega$ integral yields the multivariate Gaussian of the form
\eqref{multivar}, but now with the covariance matrix $z\Sigma/N$.
Hence, we arrive at the remarkable expression
\begin{equation}
\langle  g \rangle (r|\Sigma,N) = \int_0^\infty \chi_N^2(z) g\left(r\Big|\frac{z}{N}\Sigma\right) dz \ ,
\label{compound}
\end{equation}
which maps the whole random matrix average to an one--dimensional
average involving the $\chi^2$ distribution of $N$ degrees of freedom,
\begin{equation}
\chi_N^2(z) = \frac{1}{2^{N/2}\Gamma(N/2)} z^{N/2-1} \exp\left(-\frac{z}{2}\right) 
\label{chi2}
\end{equation}
for $z\ge 0$ and zero otherwise. The parameter $N$ can now be
interpreted as an effective number of degrees of freedom
characterizing the ensemble induced by the fluctuations of the
covariance matrices.  In fact, the Wishart distribution can be viewed
as a matrix generalized $\chi^2$ distribution. The integral
\eqref{compound} can be done in closed form, 
\begin{eqnarray}
\langle  g \rangle (r|\Sigma,N) &=& 
                \frac{1}{2^{N/2+1}\Gamma(N/2)\sqrt{\det(2\pi\Sigma/N)}} \nonumber\\
   & & \qquad \frac{\mathcal{K}_{(K-N)/2}\left(\sqrt{Nr^\dagger\Sigma^{-1}r}\right)}
                                      {\sqrt{Nr^\dagger\Sigma^{-1}r}^{(K-N)/2}} \ ,
\label{ergebnis}
\end{eqnarray}
where $\mathcal{K}_\nu$ is the modified Bessel function of the second
kind of order $\nu$.  In the data analysis below, we will find $K>N$.
Since the empirical covariance matrix $\Sigma$ is fixed, $N$ is the
only free parameter in the distribution \eqref{ergebnis}.  For large
$N$ it approaches a Gaussian. The smaller $N$, the heavier the tails,
for $N=2$ the distribution is exponential. Importantly, the returns
enter $\langle g \rangle (r|\Sigma,N)$ only via the bilinear form
$r^\dagger\Sigma^{-1}r$. This high degree of invariance is due to the
invariance of the Wishart distribution \eqref{wishart}. Such features are common to all random matrix models. 
The derivation and the result \eqref{ergebnis} extend and
generalize our previous study \cite{Munnix2011a} in which we aimed at
rough estimates for another purpose, and in which we did not carry out a data analysis.
Hence, the decisive question is, if our new result describes the data. 

\begin{figure}[htbp]
  \begin{center}
    \includegraphics[width=0.49\textwidth]{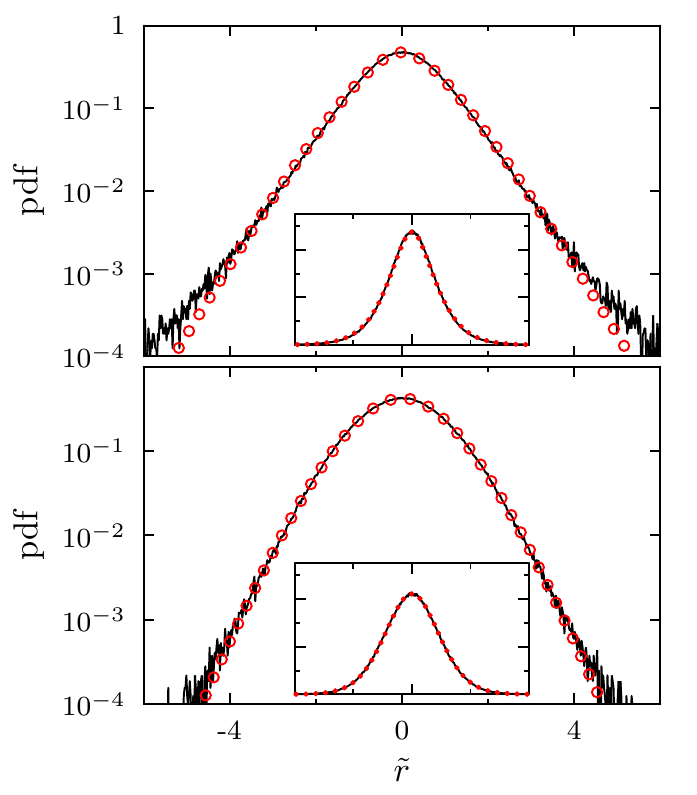}
  \end{center}
  \caption{Aggregated distribution of the rotated and scaled returns
    $\tilde{r}$ for $\Delta t =1$ (top) and $\Delta t =20$ (bottom)
    trading days. The circles  correspond to the
    distribution~(\ref{rk}).}
 \label{fig6}
\end{figure}

\begin{figure}[htbp]
  \begin{center}
    \includegraphics[width=0.42\textwidth]{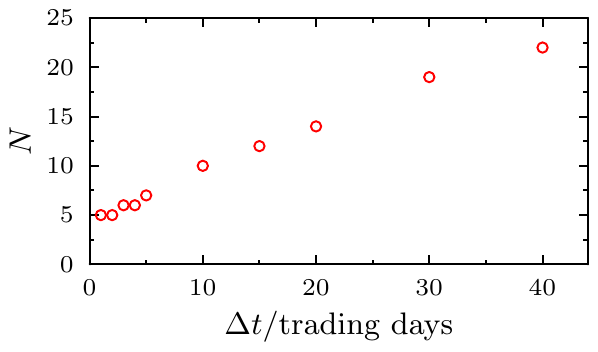}
  \end{center}
  \caption{The parameter $N$ versus the return interval $\Delta t$.}
 \label{ndep}
\end{figure}

To test this, we rotate the vector $r$ into the eigenbasis of
the covariance matrix $\Sigma$, and normalize its elements with the
eigenvalues.  Integrating out all but one of the components of the
rotated vector, which we denote $\tilde{r}$, we find
\begin{align}
\langle g \rangle(\tilde{r}|N) & = \frac { \sqrt{ 2 }^{1-N} \sqrt{N} }{ \sqrt{\pi} \Gamma(N/2) } \sqrt{N \tilde{r}^2}^{(N-1)/2} \mathcal{K}_{(N-1)/2} \left(\sqrt{N \tilde{r}^2} \right)
\label{rk}
\end{align}
This formula is compared in Fig.~\ref{fig6} with the aggregated
distributions evaluated form the SP500 dataset. We determine the
parameter $N$ with a Cramer--von Mises test~\cite{Anderson1962} considering only integer
values and find $N=5$ for daily returns and $N=14$ for $\Delta t =20$
trading days. We find a good agreement between model and data.
Importantly, the distributions have heavy tails which result from the
fluctuations of the covariances, the smaller $N$, the heavier.
For small $N$ there are deviations between theory and data in the tails.
Figure~\ref{ndep} shows that $N$ increases monotonically with the return
interval $\Delta t$.

As already mentioned, our random matrix model has invariances.  On the
other hand, Fig.~\ref{fig3} clearly shows that the empirical ensemble
of correlation matrices has inner structures and is thus not at all
invariant. Interestingly, our \textit{invariant} random matrix model can handle this. 
The effective number of degrees of
freedom $N$ also contains information about these inner structures.

The determination of the parameter $N$ by fitting is strongly
corroborated by another, independent check: The expectation value for
the bilinear form $r^\dagger\Sigma^{-1}r$ reads
\begin{equation}
\left< r^\dagger\Sigma^{-1}r \right> = 
\frac{2 \; \Gamma\left( (N+1)/2 \right) \, \Gamma\left( (K+1)/2 \right)}
{\sqrt{N} \; \Gamma\left( N/2 \right) \, \Gamma\left( K/2 \right)} \ .
\end{equation}
Comparing this result to the average value of the empirical bilinear
form yields a value of $N$ which is consistent with the fitted value.

Our formula \eqref{compound} can be viewed as an average of the
Gaussian with argument $\sqrt{Nr^\dagger\Sigma^{-1}r}$ over the
variance $z$ which is $\chi^2$ distributed. It is thus a certain
non--uniform average of Gaussians. Procedures of the type
\eqref{compound}, \textit{i.e.}, construction of a new distribution by
averaging a parameter of a distribution, are known as
\textit{compounding} \cite{dubey1970compound} or \textit{mixture}
\cite{Barndorff-Nielsen1982,Doulgeris2010} in the statistics
literature.  Starting from a compounding ansatz, the distribution
\eqref{ergebnis} was found recently in Ref.~\cite{Hohmann2010} when
studying scattering of microwaves in random potentials. In this case,
however, the argument was simply the squared intensity and not a
bilinear form.  To the best of our knowledge, we have given here, in
the context of finance, the first \textit{explanation} of such a
compounding ansatz by deriving it from fluctuating covariances.

Random matrix models are widely used in quantum chaos, many--body and
mesoscopic physics and in related fields, see Ref.~\cite{Guhr1998} for
a review. The systems in question are quantum dynamical systems
defined by a Hamiltonian. Thus, there is a fundamental difference to
be underlined when comparing to systems such as financial markets. Of
course, the latter are also dynamical systems, but of a very different
nature. In particular, there is nothing like the mean level spacing in
quantum chaotic or mesoscopic systems which sets the scale when
investigating spectral fluctuations. Thus, there cannot be the type of
universality as discussed so much in the context of random matrix
models based on a Hamiltonian. In view of this, it is even more
encouraging that our present random matrix model, which is just based
on the Gaussian Wishart distribution, leads to a quantitative
description of the heavy--tailed return distributions. Another
important caveat is in order. When measuring correlations, the effect
of noise--dressing leads to an eigenvalue density for
\textit{individual} correlation matrices which is consistent with
Wishart random matrices \cite{Laloux1999,Plerou1999a}. This is not
related to our study. We model fluctuating correlations by an
\textit{ensemble} of covariance or correlation matrices. The only
correlation matrix which we directly extract from the data is for the
entire observation period. Measurement noise is thus negligible.

In summary, we have shown that the fluctuations of correlations and
covariances induce generic properties. We uncovered them by
constructing a random matrix model which reduces the high complexity
of a whole correlated market to one single parameter that
characterizes the fluctuations. Although this finding is reminiscent
of equilibrium statistical mechanics, we emphasize once more that we
treated a substantially non--stationary system. Our model is capable
of describing the empirical non--invariant ensembles of covariance
matrices in terms of an invariant random matrix ensemble. It also
quantitatively explains how the fluctuations yield heavy tails.

We thank Sonja Barkhofen and Hans--J\"urgen St\"ockmann for fruitful
discussions on the r\^ole of the $\chi^2$ distribution and for drawing
our attention to their work on waves.

\end{document}